\documentclass[english,aps,prl, 10pt,twocolumn,superscriptaddress]{revtex4-1}
\usepackage[T2A,T1]{fontenc}
\usepackage[latin9]{inputenc}
\usepackage{color}
\usepackage{amsmath}
\usepackage{amssymb}
\usepackage{graphicx}

\makeatletter

\DeclareRobustCommand{\cyrtext}{%
  \fontencoding{T2A}\selectfont\def\encodingdefault{T2A}}
\DeclareRobustCommand{\textcyr}[1]{\leavevmode{\cyrtext #1}}
\AtBeginDocument{\DeclareFontEncoding{T2A}{}{}}

\@ifundefined{textcolor}{}
{%
 \definecolor{BLACK}{gray}{0}
 \definecolor{WHITE}{gray}{1}
 \definecolor{RED}{rgb}{1,0,0}
 \definecolor{GREEN}{rgb}{0,1,0}
 \definecolor{BLUE}{rgb}{0,0,1}
 \definecolor{CYAN}{cmyk}{1,0,0,0}
 \definecolor{MAGENTA}{cmyk}{0,1,0,0}
 \definecolor{YELLOW}{cmyk}{0,0,1,0}
}

\newcommand{\barrier}{\mbox{\tiny barrier}}
\newcommand{\switch}{\mbox{\tiny switch}}
\newcommand{\mono}{\mbox{\tiny mono}}

\newcommand{\eff}{\mbox{\tiny eff}}

\newcommand{\fs}[1]{{#1}}

\makeatother

\usepackage{babel}
\begin{document}

\preprint{This line only printed with preprint option}

\title{\textcolor{black}{
Sharp and fast: Sensors and switches based on polymer brushes with adsorption-active minority chains.
}}

\author{\textup{Leonid I. Klushin}}

\affiliation{Department of Physics, American University of Beirut, P.O. Box 11-0236,
Beirut 1107 2020, Lebanon}

\author{\textup{Alexander M. Skvortsov}}

\email{astarling@yandex.ru}

\affiliation{Chemical-Pharmaceutical Academy, Prof. Popova 14, 197022 St. Petersburg,
Russia}

\author{\textup{Alexey A. Polotsky}}

\email{alexey.polotsky@gmail.com}

\affiliation{Institute of Macromolecular Compounds of Russian Academy of Sciences,
31 Bolshoy pr., 199004 Saint-Petersburg, Russia}

\author{\textup{Shuanhu Qi}}

\affiliation{Institut f\"ur Physik, Johannes-Gutenberg Universit\"at Mainz, Staudinger
Weg 7-9, 55099 Mainz, Germany}

\author{\textup{Friederike Schmid}}

\email{friederike.schmid@uni-mainz.de}

\affiliation{Institut f\"ur Physik, Johannes-Gutenberg Universit\"at Mainz, Staudinger
Weg 7-9, 55099 Mainz, Germany}

\begin{abstract}
\textcolor{black}{
%Typically sensor switches are made from mixed polymer
%brush from hydrophilic and hydrophobic polymer species randomly grafted
%onto the same solid substrate. The switching causes by phase segregation
%and occures in wide temperature interval $\triangle T\approx30$ C
%with slow response on the timescale from minutes to hours.
We propose a design for polymer-based sensors and switches with sharp switching
transition and fast response time. The switching mechanism involves a radical change
in the conformations of adsorption-active minority chains in a brush. Such
transitions can be induced by a temperature change of only about ten degrees,
and the characteristic time of the conformational change is less than a second.
We present an analytical theory for these switches and support it by
self-consistent field calculations and Brownian dynamics simulations.
}
\end{abstract}
\maketitle

Multicomponent polymer brushes offer promising perspectives for the design of
smart responsive materials with a wide range of applications in nano- and
biotechnology \cite{CohenStuart:2010, Chen:2010}.  For example, mixed polymer
brushes comprising approximately equal amounts of hydrophobic and hydrophilic
polymers have been used to fabricate surfaces with switchable wettability
\cite{Draper:2004, Motornov:2003}. If the polymers phase separate along the
direction perpendicular to the brush, the surface properties can switch between
the properties of the two polymer species. The transition occurs on a
temperature interval of $\Delta T \approx 30 K$.  However, the response times
are relatively slow on the time scale of minutes to hours, due to the
existence of kinetically frozen metastable states with lateral nanoscale
segregation.

\begin{figure}[t]
\centerline{\includegraphics[width=7.5cm]{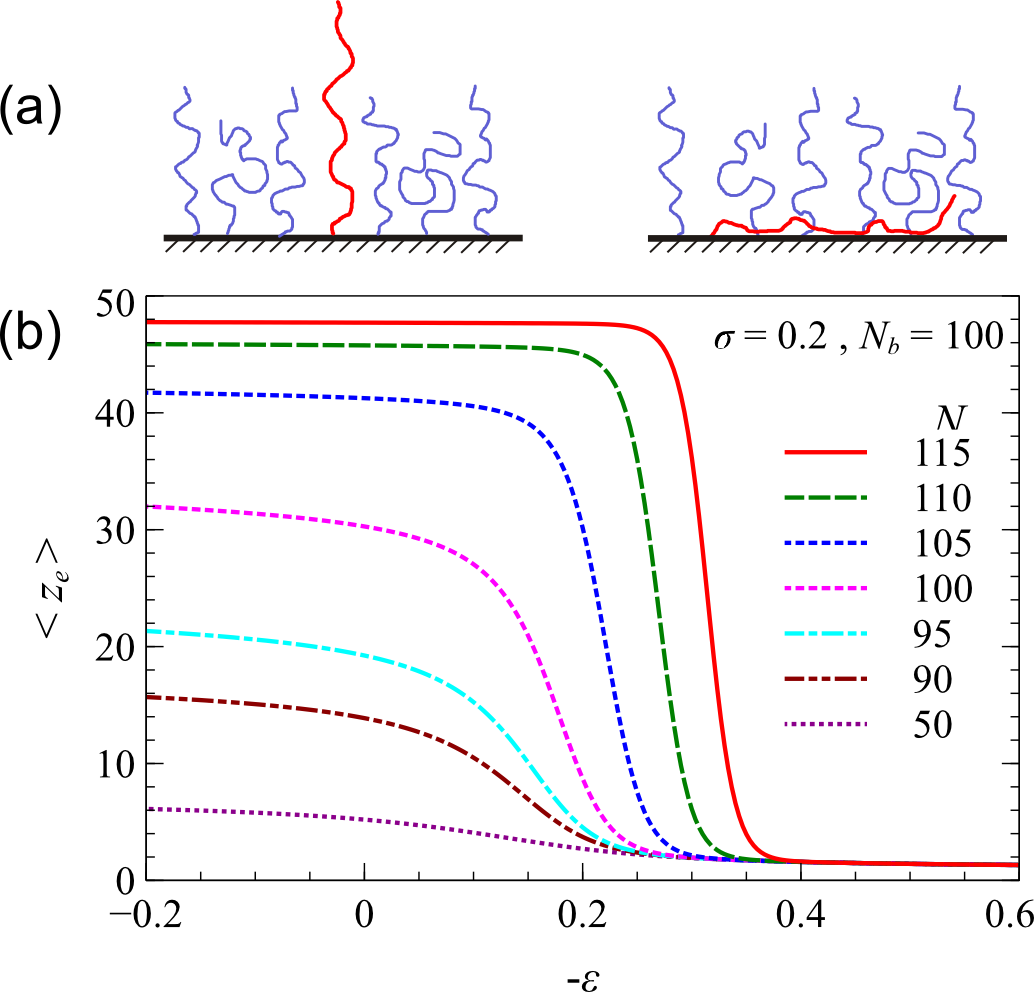}}
\vspace*{-0.2cm}
\caption{\label{fig:z_average}
(a) Cartoon of the two states of the switching minority chain.
(b) Average chain end height $\langle z_e\rangle$ vs. adsorption
energy $\varepsilon$ for different values of the minority chain length $N$
(as indicated) in a brush with chain length $N_{b}=100$ and grafting 
density $\sigma=0.2$.
}
\end{figure}

In this letter we propose a new class of brush-based switches, which rely on a
radical conformational change of adsorption-active minority chains in a brush.
The basic mechanism of the transition is illustrated in Fig.\
\ref{fig:z_average}a). Consider a brush of polymers with chain length $N_b$
containing a small amount of minority chains with length $N>N_b$, which undergo
an adsorption transition on the substrate. In the absence of the brush, the
adsorption transition (in the limit $N \to \infty$) is continuous. Due to the
interaction with the brush, it becomes first order at $N,N_b \to \infty$, and
at finite chain length the chain end distribution of the minority chain becomes
bimodal. A small change in temperature or solvent composition may thus lead to
a sharp transition from an adsorbed state, where the switch chain is completely
hidden inside the brush, to an exposed state, where the free end of the switch
chain is localized at the outer surface of the brush. If each minority chain
has an active group attached to its free end, the brush switches between two
states: one where all active groups are fully hidden inside the brush, near the
solid substrate, and one where they are exposed at the outer brush surface.
The active end-groups can serve as sensors triggering an immune-like response
or a detectable change in optical properties.  

The proposed switches possess two main advantages.  First, the transition is
sharp and can be induced by a temperature change of only about ten degrees, as
attested and utilized by polymer chromatography in mixed eluents
\cite{Abdulahad:2009}. Second, the characteristic time scale for conformational
changes is small; below, we estimate it to be well below a second. Hence the
rate of change in brush properties is limited by the rate of change in the
external conditions. We should note that a similar conformational transition
can be observed when varying the contour length of the minority chain, and this
has been studied in earlier work by theory \cite{Skvortsov:LongShort1997} and
simulations \cite{Sommer:2012,Romeis:2013}.  However, this effect cannot be
used for switching since it is almost impossible to change the contour length
in a controlled and reversible way. 

The purpose of the present letter is to provide a theoretical description for
the proposed class of sensor switches and to clarify how the main brush
parameters (the chain lengths $N_b$ and $N$, the grafting density $\sigma$)
affect the two main characteristics of the switch performance: the sharpness of
the transition from one state to the other, and the time of response to a
sudden change in the control parameters. We develop an analytical theory and
support it by self-consistent field (SCF) lattice calculations and Brownian
dynamics simulations.

\begin{figure}[t]
\includegraphics[width=8cm]{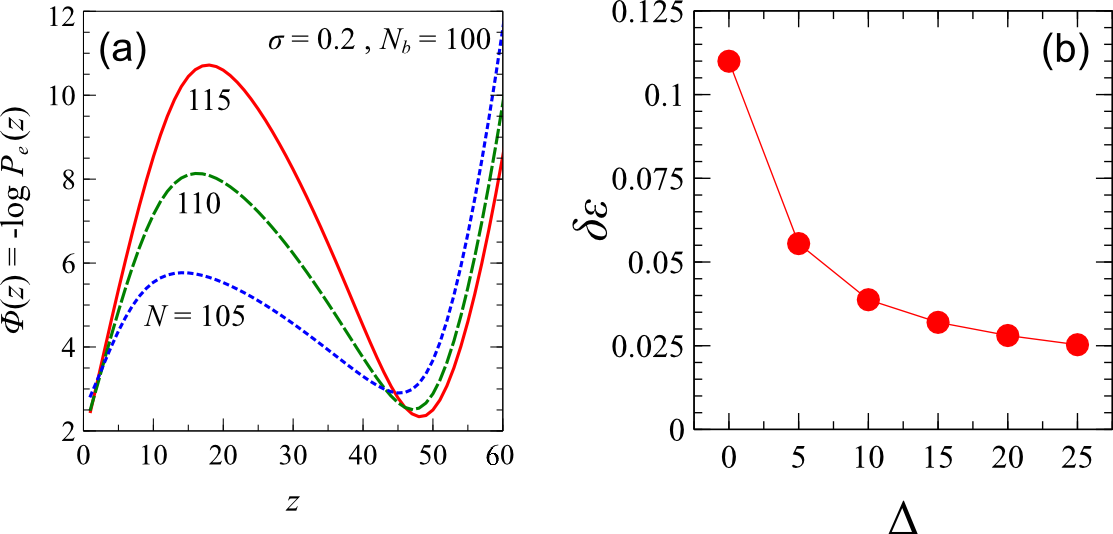} 
\vspace*{-0.2cm}
\caption{\label{fig:z_distribution} 
(a) Landau free energy (logarithm of the end distribution $P_{e}(z)$)
for switch chains of different length $N$ integrated into a brush with 
chain length $N_{b}=100$ 
and grafting density $\sigma=0.2$, at the mid-point $\varepsilon_{m}$
of the transition. Solid red line: $N=105$, $\varepsilon_{m}=0.190$;
dashed green line: $N=110$, $\varepsilon_{m}=0.246$;
dotted blue line: $N=115$, $\varepsilon_{m}=0.293$.
(b) Width of the transition between the adsorbed and desorbed chain, defined as
$\delta \varepsilon = H \cdot ({\rm d} \langle z_e \rangle/{\rm d} \varepsilon)^{-1}$
evaluated at $\varepsilon_{m}$ ($H$ is the brush height)
as a function of chain increment $\Delta=N - N_{b}$.  
}
\end{figure}

\textbf{\textit{Evidence of the sharp transition.}} 
To illustrate the switch mechanism, we first present SCF results for model
brushes of polymers with length $N_b=100$ containing a single 
adsorption-active chain of length $N$. The minority chain is subject to the
SCF brush potential created by the other chains, plus an adsorption potential:
Each monomer contact with the surface leads to an energy gain $\varepsilon$ .
In the following, all energies are given in units of the thermal
energy, $k_B T$, and all lengths in units of the segment length, $a$. Fig.\
\ref{fig:z_average}b) shows the average distance between the free end of the
adsorption-active switch chain to the solid substrate, $\langle z_e \rangle$, as
a function of $\varepsilon$ for brushes with grafting density $\sigma = 0.2$.
The curves are obtained by SCF calculations for walks on a 6-choice simple
cubic lattice.  Already a small positive increment $\Delta=N-N_{b}$ in the
length of the minority chain results in a pronounced transition from a
stretched state in which the free end of the minority chain is localized at the
outer surface of the brush to an adsorbed state with the switch chain localized
very close to the substrate. For negative or zero increment, the transition is
considerably smoother, and the desorbed state
displays large fluctuations of the chain end position, as typical for
polymer brushes \cite{Zhulina:Lai}.  In contrast, the end fluctuations for
desorbed chains with positive chain length increment $\Delta$ are much smaller,
and the peak of the distribution is located near the brush edge
\cite{Skvortsov:LongShort1997,Romeis:2013}.  This is illustrated in Fig.\
\ref{fig:z_distribution}a), which shows the logarithm of the end distribution
(the Landau free energy for chains with end position $z$) for different switch
chain length $N$ at the mid-point of the transition, i.e., the adsorption
strength $\varepsilon_{m}$ where the switch chain can be found in the adsorbed 
and the desorbed state with equal probability. The free energy has two distinct
minima, hence the two states are clearly separated. The location of the minimum
at large $z$ corresponds exactly to the most probable chain end position in 
the absence of an attractive surface potential.  The minimum at small $z$ 
refers to rather strongly adsorbed chains since the adsorption energies are 
considerably larger than the critical adsorption energy
$\varepsilon_{c}=-\ln(6/5)\simeq-0.182$ \cite{Rubin:JCP:1965}.  With increasing
$N$, the minima become separated more strongly and the transition sharpens,
i.e., the width of the transition in $\varepsilon$ decreases
(Fig.\ \ref{fig:z_distribution}b).

\textbf{\textit{Theory}}.
The theoretical description of the two competing
states of the adsorption-active minority chain of length $N$ surrounded
by the non-adsorbing chains of length $N_{b}$ is based on several
simplifying assumptions: First, we treat the inter- and intrachain
interactions in the mean-field approximation. Second, we assume that
minority chains are grafted far enough from each other so that their
mutual interaction is negligible compared to interactions of the minority
chains with the surrounding brush chains. Third, we neglect the 
effect of the change in the minority chain conformation on the surrounding
brush. Hence, the conformation of a minority chain is determined by
a fixed mean-field potential profile consisting of the repulsive contribution
determined by the brush density and a short-ranged attraction 
to the solid substrate. The minority chain itself is described by
an ideal continuum model, since intrachain excluded volume effects
are screened out considerably within the brush thickness. The Green's
function $G(z,N)$, i.e. the total statistical weight of the minority
chain grafted at the substrate ($z_{0}=0$) as a function of the free
end position, $z$, is a solution of the Edwards equation 

\begin{equation}
\frac{\partial G(z,s)}{\partial s}=\frac{1}{6}\frac{\partial^{2}G(z,s)}{\partial^{2}z}-V\left(z\right)G(z,s)\,,\label{eq:Edwars-Green}
\end{equation}
where $s$ is the number of monomers. The total potential is $V(z)=V_{b}(z)+V_{ads}(z)$,
where the mean field due to a brush of thickness $H$ is known from
analytical theory \cite{ZhulinaBorisovPriamitsyn}:
\begin{equation}
V_{b}(z)=\frac{3\pi^{2}}{8N_{b}^{2}}\left(H^{2}-z^{2}\right)\,,
\label{eq:brush_potential}
\end{equation}
and the adsorption interaction is described by an attractive pseudopotential
of strength $c$, $V_{ads}(z)=-c\delta(z)$. Close to the substrate,
$z\ll H$, the brush potential is roughly constant, 
$V_{b}(z)\simeq\frac{3\pi^{2}}{8N_{b}^{2}}H^{2}\equiv U_{0}$.
For moderate grafting densities, $\sigma\lesssim0.25$, the analytical
theory gives a simple relation, $U_{0}=\frac{3}{2}\left(\frac{\pi\sigma}{2}\right)^{2/3}$
\cite{ZhulinaBorisovPriamitsyn}. 

The adsorbed state is thus approximately described by the standard
partition function modified by the potential $U_{0}$ 
\cite{Skvortsov:MinorityAdsorption1999}:
\begin{equation}
Q_{ads}=2e^{N(-U_{0}+c^{2}/6)}\label{eq:Qads}
\end{equation}
The state with the free end exposed at the brush edge has no contacts
with the surface. Hence, the Green's function is given by the known
solution of the Edwards equation for a purely parabolic potential
and a neutral solid surface \cite{Skvortsov:LongShort1997} :
\begin{equation}
G_{ex}(z,N)=\frac{\pi}{2}\left(\frac{3}{N_{b}\sin\left(\frac{\pi N}{2N_{b}}\right)}\right)^{3/2}z\, e^{-\frac{3\pi}{4N_{b}}\cot\left(\frac{\pi N}{2N_{b}}\right)z^{2}},\label{eq:Gex}
\end{equation}

For minority chains with positive increment, $N>N_{b}$, the Green's function,
Eq.(\ref{eq:Gex}), increases monotonically with $z$ and is unbounded, since
it refers to a parabolic potential (\ref{eq:brush_potential}) extending to infinity.
% and to an infinitely extensible Gaussian chain
% FS: I don't understand why this matters here. 
In reality, the potential is cut off at $z=H$, and the
minority chain does not make excursions well beyond $z=H$. To account for this,
we truncate $G_{ex}(z,N)$ at $z=H$. If minority chains are just slightly
longer $\Delta\equiv N-N_{b}\ll N_{b}$, (which is of prime interest for this
letter) the Green's function simplifies to: 

\begin{equation}
G_{ex}(z,N)=\begin{cases}
\begin{array}{ccc}
\frac{\pi}{2}\left(\frac{3}{N_{b}}\right)^{3/2}z\, e^{\left(U_{0}\Delta\right)\left(\frac{z}{H}\right)^{2}} & , & z\leq H\\
0 & , & z>H
\end{array}\end{cases}\label{eq:Gexpo}
\end{equation}
which gives the partition function
\begin{equation}
Q_{ex}=\frac{2}{\pi\Delta}\left(3N_{b}\right)^{1/2}e^{U_{0}\Delta}\label{eq:Qexpo}
\end{equation}
assuming $e^{U_{0}\Delta}\gg1.$ Eqs.\ (\ref{eq:Qads}),(\ref{eq:Qexpo})
clearly show that in the limit of $N_{b}\rightarrow\infty,$ 
$\sigma={\rm const},$ $\frac{\Delta}{N_{b}}={\rm const}$, the free energies of 
the two states are extensive and switching becomes a classical first-order phase 
transition.  
\fs{In the desorbed state, the chain end position $z_e$  takes the mean value 
$\langle z_e \rangle \stackrel{>}{\sim} H \propto N_b$
and the amplitude of fluctuations roughly scales like
$\Sigma_e = \sqrt{\langle z_e^2 \rangle - \langle z_e \rangle^2}
\propto \sqrt{\Delta}$.  In the adsorbed state,
$\langle z_e \rangle$ and $\Sigma_e$ are of the same order (the adsorption
blob size) and close to zero. Hence the two states are well separated.}

%\textbf{Theory predictions and numerical verification}
%\textbf{\textit{Transition point}} -- 

The theory can be used to derive approximate expressions for the transition
point and other quantities of interest. The transition point is defined
by the condition $Q_{ads}(c^{*})=Q_{ex}$ . Omitting logarithmic corrections
this gives
$
%\begin{equation}
c^{*}=\left(6U_{0}{\Delta}/{N}\right)^{1/2}.
%\end{equation}
$
The connection between the pseudopotential strength $c$ and the lattice
parameter $\varepsilon$ for 6-choice walks is known \cite{Gorbunov:2001}
and can be linearized close to the critical adsorption point as 
$c=5\left(\varepsilon_{c}-\varepsilon\right)$.
Thus the prediction for the transition point in the lattice model
reads
\begin{equation}
\left(\varepsilon_{c}-\varepsilon^{*}\right)
=\frac{3}{5}\left(\frac{\pi\sigma}{2}\right)^{1/3}\left(\frac{\Delta}{N}\right)^{1/2}
\end{equation}

\begin{figure}[t]
\includegraphics[width=8cm]{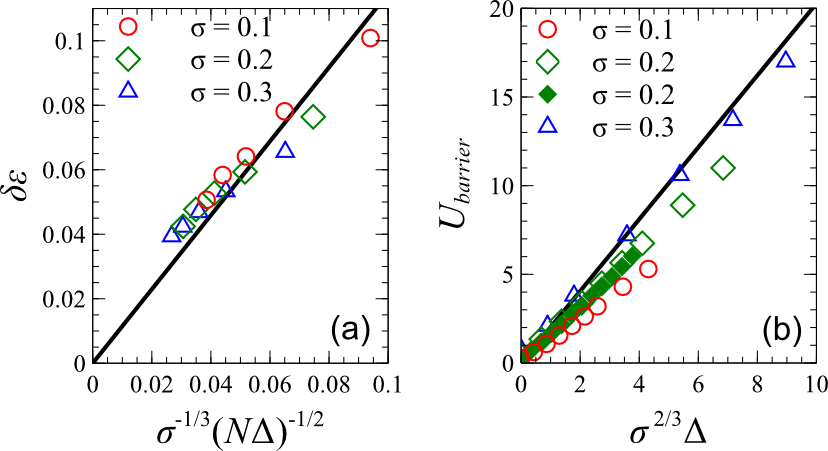}
\vspace*{-0.2cm}
\caption{\label{fig:scaling}Scaling plots for the width of the transition
(a) and the barrier height for the adsorbing minority chain (b) in brushes
with different grafting densities, plotted according
to the theoretical predictions of of Eqs.\ (\protect\ref{eq:de_scaling}) 
(a) and (\protect\ref{eq:ub_scaling}). Here $N$ is the minority chain
length and $\Delta = N - N_b$ the chain length difference between switch
chains and brush chains ($N_b$ is the brush chain length). Open symbols 
represent SCF data for $N_b = 100$ and $\sigma$ = 0.1,0.2, 0.3; filled
diamonds in panel (b) show data from single-chain Langevin dynamics
simulations, see text below.
}
\end{figure}

The sharpness of the transition, characterized by the inverse slope
$\delta \varepsilon$ of $\langle z_e \rangle$ at the mid-point of
the transition (see Fig.\ \ref{fig:z_distribution}b), can be estimated 
from a standard recipe for the two-state model \cite{BinderLandau}
$\delta\varepsilon=4\Big/\left({{\rm d} \ln Q_{ads}}/{{\rm d}\varepsilon}
\right)_{\varepsilon=\varepsilon*}$,
which combined with Eq. (\ref{eq:Qads}) results in 
\begin{equation}
\label{eq:de_scaling}
\delta\varepsilon=\frac{4}{3}\left(\frac{\pi\sigma}{2}\right)^{-1/3}\left(N\Delta\right)^{-1/2}.
\end{equation}
The SCF data are in accordance with this analytical scaling prediction, 
see Fig.\ \ref{fig:scaling}a).

Finally, we consider the height of the barrier separating the adsorbed and
exposed state, which is one important determinant of the switching time. The
sawtooth shape of the free energy potential $\Phi(z)$ in Fig.\
\ref{fig:z_distribution}a) suggests that the transition states mostly
correspond to squeezed desorbed chains.  Hence we estimate the barrier height
at the transition point, $U_{\barrier}$ , by the change in the Green's function
of the exposed state,
$U_{\barrier}=\ln\frac{G_{ex}(z=H)}{G_{ex}(z=z_{\barrier})}$.  From
Eq.(\ref{eq:Gexpo}), it follows that up to a numerical coefficient close to 1,
\begin{equation}
\label{eq:ub_scaling}
U_{\barrier}\simeq U_{0}\Delta=\frac{3}{2}\left(\frac{\pi\sigma}{2}\right)^{2/3}\Delta\,.
\end{equation}
Fig.\ \ref{fig:scaling}b) shows that the SCF data again agree reasonably well
with the analytical scaling prediction \fs{at sufficiently high grafting
density}. Here the numerical data were obtained
for the adsorption strength where the two minima of $\Phi(z)$ have equal 
depth (which is slightly different from the mid-point $\varepsilon_m$).

\textbf{\textit {Switching times.}} 
The expression for the barrier height, Eq.\ \ref{eq:ub_scaling}, can be used 
to derive a simple estimate for the switching time \cite{PaulBaschnagel}
\begin{equation}
\label{eq:switch}
\tau_{\switch}=2 \pi \tau_{0} \exp(U_{\barrier})
\end{equation}
where the time scale $\tau_0$ is related to the rate at which the chain 
attempts to overcome the barrier. A robust estimate from above is obtained by
taking the relaxation time in the absence of the barrier, i.e., the relaxation
time for switch chains with length $N=N_b$ at $\varepsilon=0$.
The single-chain relaxation time in the
presence of the mean field created by the surrounding brush 
scales as $\tau_{0}\sim N_{b}^{3}\sigma^{2/3}\tau_{\mono}$ 
\cite{Klushin:1991, Murat:1989}
where $\tau_{\mono}$ is the characteristic monomer relaxation time which is
typically on the order of $10^{-9}s$ for flexible chains. 

\begin{figure}[t]
\includegraphics[width=8cm]{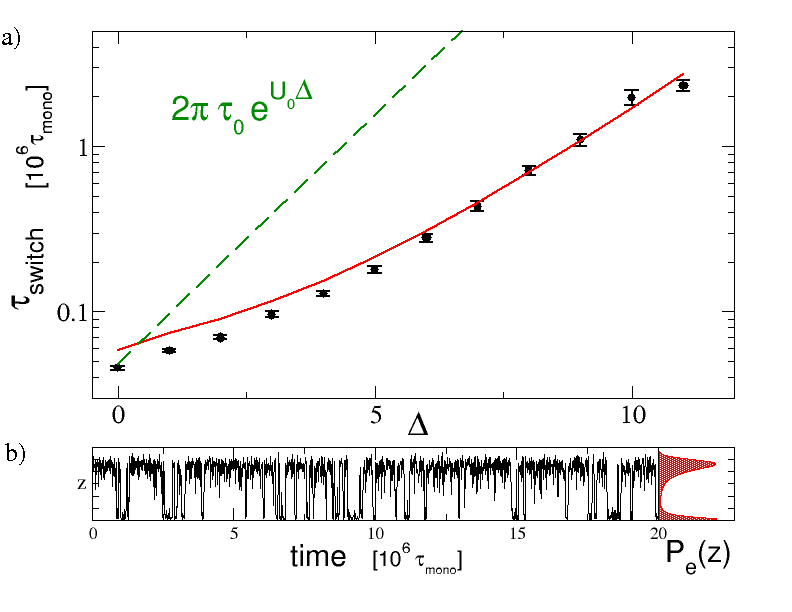}
\vspace*{-0.2cm}
\caption{\label{fig:tau} 
(a) Mean switching times for the switch chain in a brush with
brush chain length $N_b=100$ and grafting density $\sigma=0.2$
at the adsorption transition (see text)
as a function of the length increment $\Delta = N-N_b$.
Symbols: data from Brownian dynamics simulations; 
dashed green line: Arrhenius estimate (Eq.\ (\protect\ref{eq:switch}));
solid red line: refined estimate based on Kramers' theory
(Eq.\ (\protect\ref{eq:switch2})) \fs{with 
$N_{\eff}= N - N_b/3$}.
(b) Time evolution of the end monomer position $z$ and 
corresponding end monomer distribution $P_e(z)$ 
for $\Delta = 6$.
}
\end{figure}

%The Arrhenius-type expression for the switching time, Eq.\ (\ref{eq:switch}),
%is based on the picture of a single Brownian degree of freedom (the chain end
%position) trying to overcome a high energy barrier $U_{\barrier}$. In reality,
%the barrier height is only a few $k_B T$ (see Fig.\ \ref{fig:z_distribution}a)
%and the transition is a collective event that involves several chain modes. 
In order to obtain a more realistic description of the switching dynamics, we have
carried out overdamped Brownian dynamics simulations of Gaussian switch chains
in the same SCF brush potential used for the previous calculations (at 
$N_{b}=100$, $\sigma=0.2$), plus an exponentially
decaying adsorption potential with adsorption strength chosen such that the end
monomer distribution, $P_e(z)$, has equal height in the adsorbed and desorbed
state. The static properties of the Gaussian chains are almost identical to
those of the corresponding lattice chains. For example, the barrier height
derived from the chain end distribution $P_e(z)$ distribution is the same (see
Fig.\ \ref{fig:scaling}b).  The mean switching time between states (averaged
over both switching directions) is shown in Fig.\ \ref{fig:tau}a)  as a
function of the chain length increment $\Delta$ and compared with Eq.\
(\ref{eq:switch}). Here the single-chain relaxation time in the brush, $\tau_0$,
was determined from independent simulations at $\Delta=0$ and $\varepsilon = 0$
(giving $\tau_0 = 7,7 \cdot 10^3 \tau_{\mono}$). Eq.\ (\ref{eq:switch}) strongly
overrates the switching time already for small increments $\Delta$, and the
deviations become worse for larger $\Delta$. A better estimate is obtained if
one accounts for the full shape of the free energy profile according to
Kramers' theory \cite{PaulBaschnagel}
\begin{equation}
\label{eq:switch2}
\tau_{\switch} = \frac{1}{2} N_{\eff} \tau_{\mono}
\int_{z_{ads}}^{z_{ex}} {\rm d} z \int_0^\infty {\rm d} z' \:
e^{\Phi(z)-\Phi(z')},
\end{equation}
where $z_{ads}$ and $z_{ex}$ are the positions of the two minima of $\Phi(z)$.
Here $N_{\eff}$ is an adjustable parameter denoting the effective
number of monomers that participate in the switching process. $N_{\eff}$
is smaller than $N$ due to the fact that the monomers close to the
grafted end remain close to the surface even in the exposed state. 
\fs{The simulation data can be fitted reasonably well with
$N_{\eff} \approx N - N_b/3$ (Fig.\ \ref{fig:tau}).}

According to the simulation results shown in Fig.\ \ref{fig:tau}, a switch based 
on the proposed mechanism could have switch times in the range of 
$10^6 \tau_{\mono}$, i.e., milliseconds. In real brushes, the coupling
of the minority chain with the brush will slow down the relaxation 
times \cite{BinderBrushDynamic}. \fs{On the other hand, hydrodynamic
interactions, which have also been neglected in the present work, 
tend to facilitate cooperative motions and will likely accelerate
the switching dynamics.}
%Extensive molecular
%dynamics simulations of chain dynamics in brushes \cite{BinderBrushDynamic}
%suggest that due to this coupling, the expected scaling of the
%relaxation time $\tau_{\brush} \sim N_b^{3} \tau_{\mono}$ is replaced
%by an apparent scaling $\tau_{\brush} \sim N_b^{3.7} \tau_{\mono}$.
%This could be used to estimate a correction factor. 
%note that since the switch is supposed to respond to changes in the
%environment, the actual switching process will not happen exactly
%at the transition point but rather well beyond it, so that the barrier
%will be correspondingly suppressed.
%
%FS: I dropped this part because we argue higher up that the switching
% time is so fast that the limiting factor is the change of environment.
%

In sum, we have presented a mechanism that could be used to design smart
switchable surfaces with very short response times.  The switching times are
much faster than in conventional stimulus-responsive surfaces based on mixed
brushes, because they do not involve a cooperative reorganization of an entire
polymer layer.  The theory developed above indicates that the two main
requirements for an optimal switching process, i.e., fast response times and
high sensitivity to changes in the environment, are at conflict with each
other. The width of the transition decreases with increasing chain length
increment $\Delta$ of the switch chain (Fig.\ \ref{fig:z_distribution}b)),
which increases the sensitivity, but at the same time, the switching time
increases (Fig.\ \ref{fig:tau}a).  Nevertheless, we can identify a regime,
$\Delta/N_b \sim 10 \%$, where the transition is sharp while the switching times
are still well in the subsecond regime. 

\fs{The mechanism relies on the fabrication of dense brushes in the strongly 
stretched limit. Such brushes are typically not monodisperse. Nevertheless,
the switching mechanism should still be effective.
Polydisperse brushes also create a potential barrier between the adsorbed 
and desorbed state of the switch chain. Chain end 
fluctuations of brush chains are reduced in polydisperse brushes
\cite{Klushin:1992, Vos:2009}, hence the switching transition might even 
become sharper in the regime  of small $\Delta$ where the lengths
of switch chain and brush chains are comparable.
}

\textit{Acknowledgements}

Financial support by the Deutsche Forschungsgemeinschaft (Grant SCHM
985/13-1, and 13-03-91331-NNIO-\textcyr{\char224}) is gratefully acknowledged.
We thank K. Binder for helpful discussions. Simulations have been carried
out on the compute cluster Mogon at JGU Mainz.

\bibliography{Adsorption}

\end{document}